\documentclass[journal=jpcbfk]{achemso}
\setkeys{acs}{articletitle=true}
\usepackage[version=3]{mhchem}
\usepackage{graphicx}
\usepackage{amsmath}
\usepackage{amssymb} 
\usepackage{mhchem}

\author{Matheus Girotto}
\affiliation{Instituto de F\'isica, Universidade Federal do Rio Grande do Sul, Caixa Postal 15051, CEP 91501-970, Porto Alegre, RS, Brazil}

\author{Alexandre P. dos Santos}
\affiliation{Instituto de F\'isica, Universidade Federal do Rio Grande do Sul, Caixa Postal 15051, CEP 91501-970, Porto Alegre, RS, Brazil}

\author{Yan Levin}
\email{levin@if.ufrgs.br}
\affiliation{Instituto de F\'isica, Universidade Federal do Rio Grande do Sul, Caixa Postal 15051, CEP 91501-970, Porto Alegre, RS, Brazil}

\title{Interaction of Charged Colloidal Particles at the Air-Water Interface}

\begin{document}

\begin{abstract}

We study, using Monte Carlo simulations, the interaction between charged colloidal particles confined to 
the air-water interface. The dependence of  
force on ionic strength and counterion valence is explored. 
For 1:1 electrolyte, we find that the electrostatic interaction at the interface
is very close to the one observed in the bulk.  
On the other hand, for  salts with multivalent counterions,
an interface produces an enhanced attraction between like charged colloids. 
Finally, we explore the effect of induced surface charge at the air-water interface 
on the interaction between colloidal particles.

\end{abstract}

\maketitle

\section{Introduction}

Presence of colloidal particles at a fluid-fluid interface can lower significantly the interfacial energy, 
giving them an amphiphilic character. 
The adsorption energies can be many thousands of $k_B T$'s even for relatively small colloidal particles of
only a few hundred \AA's, making the colloidal adsorption an irreversible process~\cite{Hurd85}.  
To help the dispersion in water,
colloidal particles are often synthesized with ionic groups at their surface.
When in water, these groups dissociate leading to an effective surface charge
which helps to stabilize a suspension against flocculation and precipitation.
The dissociation of surface groups is favored by
the entropic gain of the counterions release, and is opposed by 
the electrostatic
self-energy penalty.  
Adsorption of charged colloidal particles to the air-water
interface prevents the dissociation of charged groups exposed to the low dielectric environment 
since this leads to a very large electrostatic free energy penalty.

\begin{figure}[h]
\vspace{0.2cm}
\includegraphics[width=6.cm]{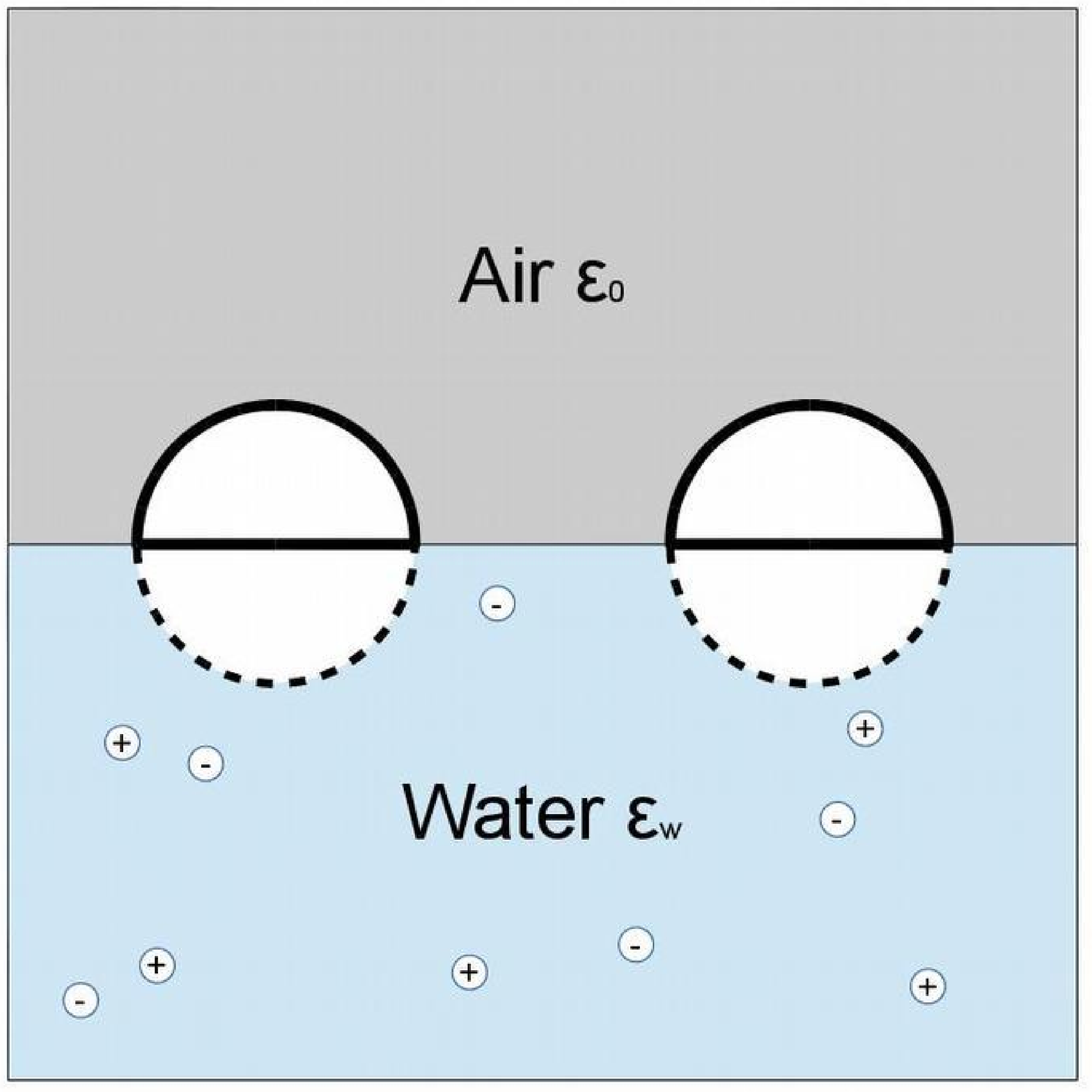}\vspace{0.2cm}
\caption{Colloidal particles at an air-electrolyte solution interface. Only colloidal hemisphere which is hydrated is charged.}
\label{fig6}
\end{figure}

Colloidal particles confined to an electrolyte-air interface exhibit many interesting properties 
which have stimulated a number of theoretical and experimental works~\cite{We02,Fo04,Die05,DoDi05,Fo05,Yo05,Ru05,To05,To06,Die07,DoOe08,FrOe11,Gra11,Die14}.
The first direct observation of colloidal bi-dimensional structure was made by Pieranski more than 3 decades ago~\cite{Pi80}. Pieranski showed that the asymptotic electrostatic potential between two colloidal particles, confined to an interface, decays as $1/r^3$.  This repulsive potential has a suggestive dipole-dipole-like form, which results in formation of two dimensional triangular lattice by the adsorbed colloidal particles.  Using linearized Poisson-Boltzmann (PB) theory, Hurd calculated the interaction potential between two point-particles confined to  the air-electrolyte solution~\cite{Hurd85} interface. The interaction potential for large separation was found to have precisely a dipole-dipole-like form observed in Pieranski's experiments~\cite{Hurd85,Pau00,Bu02,Ve08,OeDi08,Oe08}. 
 Hurd's calculations based on the linearized PB equation are sufficient to study the interaction potential at large separations for suspensions in a symmetric 1:1 electrolyte.  However, at short separations, when the electrostatic potential is large, linearization of PB equation can not be justified.  
 This requires a complicated numerical solution or some form of Derjaguin-like approximation based on the 
 non-linear solution of PB equation in a 
 planar geometry.  The situation becomes even more complicated if suspension contains multivalent counterions, in which case strong electrostatic correlations completely invalidate the use of PB theory~\cite{Le02}.
 In this paper, we will explore the interaction between two colloidal particles confined to a dielectric air-electrolyte interface using Monte Carlo~(MC) simulations.  In this respect, our work should provide a benchmark against which future analytical and numerical approximations can be tested.  The system studied is depicted in Fig.~\ref{fig6}. In the next section, we will present our model and discuss the MC method used to perform the simulations. 

\section{Model and Monte Carlo Simulations}

There are a number of different techniques developed in the literature which can be used to simulate systems with  long-range interactions. The difficulty of studying these systems is that there is no characteristic distance at which the interaction potential can be cutoff.  This prevents us from using the usual periodic boundary conditions.  Instead, to explore the thermodynamic limit, one must 
create an infinite set of periodic replicas of the system.  The particles then  interact not only with the other particles of the simulation cell, but also with the images
in all the replicas.  To efficiently sum over the replicas one can use a
3D Ewald Summation~\cite{Fr02} method. This technique is well suited to study bulk suspensions.  Unfortunately it  is not appropriate to study interfacial geometry in which a system should be replicated only in two out of three dimensions. To overcome this difficulty, there have been developed a number of simulation algorithms specifically adopted to the slab geometry~\cite{Pe79,Sp97,Ha98,Rh89,Ra89,Kl92,Ba89,Cl97,Ad97}. The 2D Ewald Summation method developed by Leeuw and Perran~\cite{Pe79} is the most accurate, but is computationally very expensive~\cite{Ad97,Cl97,Sp97}. A way to
overcome the limitations of 2D Ewald Summation  is to use a 3D Ewald Summation technique with an asymmetric  
simulation cell. The aperiodic dimension of this cell must be made to include a sufficiently large empty region, free of any charged particles. In this way, the particles in the replicas in the aperiodic dimension~(lets call it $z$ direction) will interact very weakly across the empty region~\cite{Pa96}. In order to use this technique a surface term must be added to the total energy of the system~\cite{Sm81,Ber99}, to account for the conditional convergence of the sum over the replicas. This term depends on the geometry of the system, the total electric dipole moment in the $z$ direction, and the dielectric constant of the surrounding medium~\cite{Fr02,St97,Br98,Sc94,Sc95}.

Our simulation cell has the volume $V=L^3$, with $L=200~$\AA. The electrolyte is confined in 
the region $-L/2 < x < L/2, -L/2 < y < L/2, -L/2 < z < 0$, and air in $0<z<L/2$. 
Water is treated as a continuum of dielectric constant $\epsilon_w=80\epsilon_0$, where $\epsilon_0$ is the dielectric constant of vacuum. The Bjerrum length, defined as $\lambda_B=q^2/\epsilon_w k_B T$, is  $7.2~$\AA, where $q$ is the proton charge. This value is appropriate for water at room temperature. The adsorbed colloidal particles have radius of $20~$\AA\ and charge $Z=-20q$, distributed uniformly over the hemisphere that is hydrated. This surface charge density corresponds to  $\sigma\approx-0.13~C/m^2$, close to experimental values. The hemispheres exposed to the low  dielectric environment remain uncharged.   
The salt concentration is varied between $25$ and $125~$mM, in order to explore its influence on the interaction between the colloidal particles. 
The effect of counterion valence  is explored by changing between 1:1, 2:1, and 3:1 electrolyte. 
The radius of all monovalent ions is set to $2~$\AA, and of multivalent counterions to $3~$\AA.  To account for 
colloidal surface charge, we place $98$ uniformly spaced point charges -- each of charge Z/98 -- along the surface .

The total electrostatic energy of a charge neutral system containing $N$ ions of charges $q_i$, in aqueous medium, near a dielectric interface was calculated in the Ref.~\cite{DoLe14} and  Ref.~\cite{DoLe15}. To efficiently sum over all the replicas the electrostatic potential is split into long range and short range contributions. The electrostatic energy
can then be written as
\begin{equation}
U=U_S+U_L+U_{self}+U_{cor} \ .
\end{equation}
The short range electrostatic energy is $U_S=(1/2)\sum_{i=1}^{N}q_i\phi_i^S({\pmb r}_i)$, where $\phi_i^S({\pmb r})$ is,
\begin{equation}\label{phi_S3}
\phi_i^S({\pmb r})=\sum_{j=1}^{N}{}^{'} q_j\frac{\text{erfc}{(\kappa_e |{\pmb
r}-{\pmb r}_j|)}}{\epsilon_w |{\pmb r}-{\pmb r}_j|} + \sum_{j=1}^{N}\gamma q_j
\frac{\text{erfc}{(\kappa_e |{\pmb r}-{\pmb r}'_j|)}}{\epsilon_w |{\pmb r}-{\pmb
r}'_j|} \ ,
\end{equation}
where ${\pmb r}_j$ is the position of charge $q_j$ and ${\pmb r}'_j={\pmb r}_j-2z_j \hat{\pmb z}$ is the position of the image charge $\gamma q_j$. The prime on the summation means that $j\neq i$. The dumping parameter $\kappa_e$ is set to $\kappa_e=4/L$, while $\gamma=(\epsilon_w-\epsilon_a)/(\epsilon_w+\epsilon_a)$; $\epsilon_w$ and $\epsilon_a$ are the dielectric constants of water and the dielectric material, respectively. The self-energy contribution is
\begin{equation}
U_{self}=-\frac{\kappa_e}{\epsilon_w\sqrt{\pi}}\sum_{i=1}^{N}q_i^2 \ .
\end{equation}
The long range electrostatic energy is
\begin{eqnarray}\label{U_long}
U_L = \sum_{{\pmb k}}\frac{2\pi}{\epsilon_w V |{\pmb k}|^2}
\text{exp}(-\frac{|{\pmb k}|^2}{4\kappa_e^2}) \times \nonumber \\
\left[A({\pmb k})^2 + B({\pmb k})^2 + A({\pmb k}) C({\pmb k}) + B({\pmb k}) D({\pmb k})\right] \ ,
\end{eqnarray}
where
\begin{eqnarray}
A({\pmb k})= \sum_{i=1}^{N}q_i \cos{({\pmb k}\cdot {\pmb r}_i)} \nonumber \\
B({\pmb k})=- \sum_{i=1}^{N}q_i \sin{({\pmb k}\cdot {\pmb r}_i)} \nonumber \\
C({\pmb k})= \sum_{i=1}^{N}\gamma q_i \cos{({\pmb k}\cdot {\pmb r}'_i)} \nonumber \\
D({\pmb k})=- \sum_{i=1}^{N} \gamma q_i \sin{({\pmb k}\cdot {\pmb r}'_i)} \nonumber \ .
\end{eqnarray}
These functions are easily updated for each new configuration in a MC simulation. The number of vectors ${\pmb k}$'s defined as ${\pmb k}=(2\pi n_x/L,2\pi n_y/L,2\pi n_z/L)$, where $n's$ are integers, is set to around $700$ in order to achieve fast convergence. Yeh and Berkowitz~\cite{Ber99} found that the regular 3D Ewald Summation method with an energy correction can reproduce the same results as the 2D Ewald Summation method, with a significant gain in performance. Taking into account the dielectric discontinuity and the induced image charges, the energy correction for the slab geometry is
\begin{equation}
U_{cor}=\frac{2\pi}{\epsilon_w V} M_z^2 (1-\gamma) \ ,
\end{equation}
where $M_z=\sum_{i=1}^{N}q_i z_i$ is the total electric dipolar momentum in the $\hat{z}$ direction~\cite{DoLe14}.

To perform MC simulations we use regular Metropolis algorithm with $10^5$ MC steps to achieve equilibrium, while the force averages are performed with $10^5$ uncorrelated samples, each sample obtained at an interval of $100$ movements per particle, after the equilibrium has been achieved. During the equilibration, we adjusted the length of the particle displacement to achieve an acceptance of trial moves near $50\%$. 
We then calculate the force between two colloidal particles confined
to an interface.  The total force contains both electrostatic and entropic contributions. To calculate the mean electrostatic force we use the method
of virtual displacement in which one of colloidal particles is moved
while the other colloidal particle and all the ions remain fixed, which implies that
\begin{equation}
\langle {\pmb F_{el}} \rangle =\sum_i \langle-\nabla_{{\pmb r}_i}U({\pmb r}_1,...,{\pmb r}_N) \rangle
\end{equation}
In the above expression the sum runs over the point particles which make up the colloidal surface charge. 

The entropic force that arises from the transfer of momentum during the collisions between colloidal particles and ions can be calculated using the method introduced by Wu~{\it et al.}~\cite{WuBr99} It consists of performing a small virtual displacement of colloidal particles along the line joining their centers -- while all the microions remain fixed -- and counting the number of resulting  virtual overlaps between colloidal particles and the microions. The entropic force can then be expressed as,
\begin{equation}
\beta F_{en}=\frac{<N^c>-<N^f>}{2\Delta R} ,
\end{equation}
where $N^c$ is the number of virtual overlaps between the colloidal particles with the microions after a small displacement $\Delta R=0.9~$\AA\ that brings colloidal particles closer together (superscript $c$ stands for closer) and $N_f$ is the number of overlaps of  colloidal particles with the microions after a displacement $\Delta R$ that moves the two colloidal particles farther apart (superscript $f$ stands for farther)~\cite{Le12}. 


\section{Results}

\begin{figure}[h]
\vspace{0.2cm}
\includegraphics[width=7.cm]{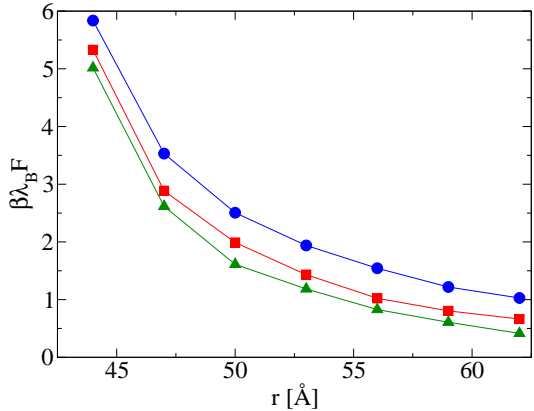}\vspace{0.2cm}
\caption{Force between two colloidal particles trapped at the  
air-water interface, for various 1:1 salt concentrations. Circles represent simulations with $25~$mM concentration of salt, squares  with $75~$mM, and triangles  with $125~$mM of salt. The inverse Debye's lenght is $\kappa^{-1} = 0.05\AA^{-1}$ for circles, $\kappa^{-1} = 0.09\AA^{-1}$ for squares and $\kappa^{-1}=0.12\AA^{-1}$ for triangles. The lines are guides to the eyes. Image effects are neglected.}
\label{fig1}
\end{figure}

We start by study the dependence of the force 
between two colloidal particles trapped at the air-water interface 
on the 
concentration of 1:1 electrolyte. 
We first neglect the dielectric discontinuity
across the interface by setting the dielectric constant of air to $\epsilon_a=\epsilon_w$, so that $\gamma=0$.
Later on we will explore the effect of the surface charge induced at the dielectric air-water interface by the microions and colloids, 
by including image charges 
in the calculation of the total force.  
In this case we will set the dielectric constant of air to $\epsilon_a=\epsilon_0$, resulting in $\gamma\approx 0.975$.
In the present paper we will neglect the image charges induced inside
the colloidal particles~\cite{DoBa11,BaDo11}  -- this is equivalent to making the two 
hemispheres of colloidal particles exposed to air and water 
to be composed of 
``air-like" and ``water-like"
dielectric materials, respectively.
In Fig.~\ref{fig1}, we plot the force vs distance curve for three different salt concentrations.

As expected, when the concentration of 1:1 electrolyte increases, the force between colloidal particles decreases. This result is the same as for  colloids in the bulk electrolyte~\cite{Le02} which are well described by the DLVO
theory.
\begin{figure}[h]
\vspace{0.2cm}
\includegraphics[width=7.cm]{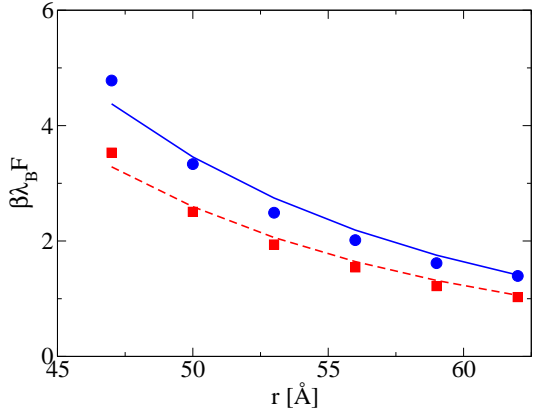}\vspace{0.2cm}
\caption{Force between two colloidal particles trapped at the air-water interface. Lines are the result of the  DLVO theory with adjusted effective charge, 
while symbols represent MC simulations. Circles are for colloidal charge $-30q$, while squares are for charge $-20q$. The inverse Debye's lengths of simulations are  $\kappa^{-1} = 0.05\AA^{-1}$. This charge is distributed uniformly over the hydrated hemisphere. 
The monovalent salt concentration is $25~$mM. Image charges are not included.}
\label{fig2}
\end{figure}
We can now see if the force between two colloidal particles at
an interface can also be described by the DLVO expression~\cite{Le02}:
\begin{equation}\label{U}
F = \dfrac{(Z_{eff})^2\theta^2(\kappa a)}{\epsilon_w} e^{-\kappa r}(\dfrac{\kappa}{r}+\dfrac{1}{r^2}) \ ,
\end{equation} 
where $Z_{eff}$ is the colloidal effective charge, $a$ is the colloidal radius, $\kappa=\sqrt{8\pi\lambda_B \rho_S}$ is the inverse Debye length, $\rho_S$ is the salt concentration, $r$ is the separation distance between particles, and $\theta(x)$ is a function given by
\begin{equation}\label{U}
\theta (x)=\frac{e^{x}}{1+x} \ .
\end{equation}

\begin{figure}[h]
\vspace{0.2cm}
\includegraphics[width=7.cm]{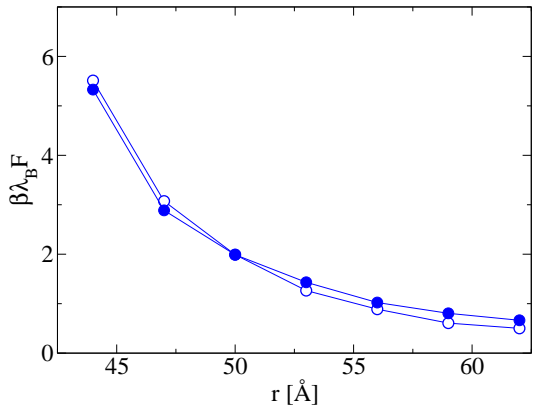}\vspace{0.2cm}
\caption{Force between two colloids. Empty symbols represent MC data for colloids in the bulk, while full ones for colloids trapped at the air-water interface. The data is for salt at $75~$mM. The inverse Debye's lengths of simulations are $\kappa^{-1} = 0.09\AA^{-1}$.  The lines are guides to the eyes. Image charges are not considered.}
\label{fig3}
\end{figure}
\begin{figure}[t]
\vspace{0.2cm}
\includegraphics[width=6.cm]{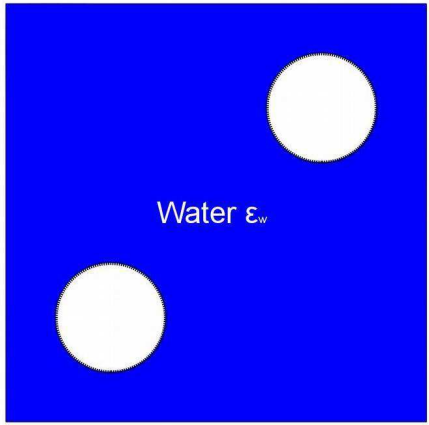}\vspace{0.2cm}
\caption{Two colloidal particles with charge $Z_{bulk}=-20q$ inside bulk electrolyte.}
\label{fig7}
\end{figure}

Fig.~\ref{fig2} shows that it is possible to  account for the MC data using DLVO theory if 
the effective charge $Z_{eff}$  is adjusted to $Z_{eff}\approx  -15.6q$ for colloidal particles of bare charge $Z=-20q$; and $Z_{eff}\approx -18q$ for colloidal particles with bare charge $Z=-30q$. 
Surprisingly, these effective charges are very close to the effective
charge of colloidal particles in the bulk which can be calculated using the Alexander's prescription~\cite{AlCh84,DoDi09}: $Z_{eff}\approx-15.9q$ and $Z_{eff}\approx-19.8q$, for colloids with bare charge $Z=-20q$ and $Z=-30q$, respectively.  This agreement is quite surprising considering that
for colloidal particles at the interface the charge is distributed only over the hydrated hemisphere, while for bulk colloids the 
{\it same} total charge $Z$ is
uniformly distributed over the whole surface. In Fig.~\ref{fig3} we present an explicit comparison of the interaction force obtained using MC simulations for two colloidal particles confined to an interface and two colloidal particles inside a bulk 1:1 electrolyte, see Fig.~\ref{fig7}. The bare charge of colloidal particles of both systems are the same.  The two forces are practically identical, see Fig.~\ref{fig7}.

\begin{figure}[h]
\vspace{0.2cm}
\includegraphics[width=7.cm]{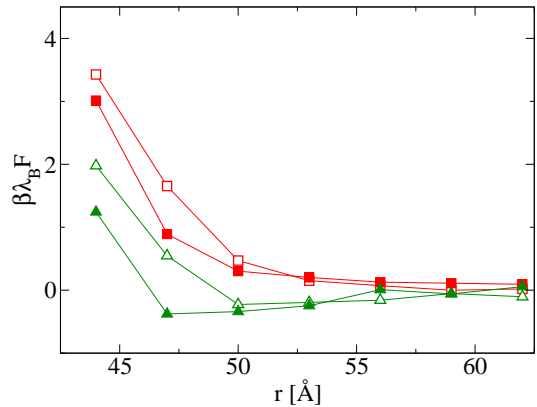}\vspace{0.2cm}
\caption{Force between two colloidal particles. Empty symbols represent the data for colloids in the bulk, while full ones for colloidal particles trapped at the water-air interface. Squares are for 2:1 electrolyte, while triangles for 3:1 electrolyte. The inverse Debye's lengths of simulations are $\kappa^{-1} = 0.11\AA^{-1}$ for squares and $\kappa^{-1} = 0.13\AA^{-1}$ for triangles.  The salt concentration is $75~$mM. The lines are guides to the eyes. Image charges are not included.}
\label{fig4}
\end{figure}

The agreement between interfacial and bulk forces does not extend to 2:1 and 3:1 electrolytes, see Fig.~\ref{fig4}. In these cases, the force in the bulk is always more repulsive than at the interface.  For 3:1 electrolyte the force at intermediate separation between the colloidal particles becomes attractive (negative).  Attraction also appears in the bulk suspensions, but is significantly weaker than at the interface.  

\begin{figure}[h]
\vspace{0.2cm}
\includegraphics[width=7.cm]{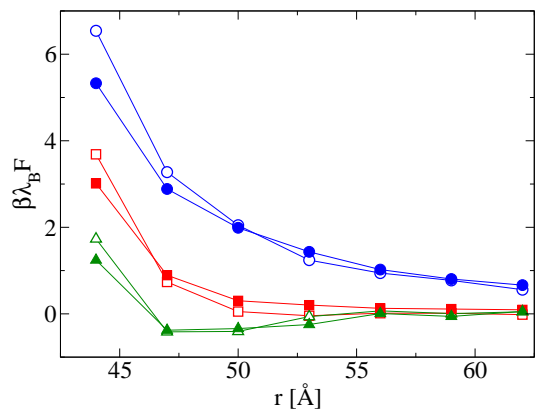}\vspace{0.2cm}
\caption{Comparison between the interaction force between two colloidal particles trapped at the air-water interface calculated with and without image charges. The full symbols represent simulations without image charges while empty symbols represent simulations with image charges. Circles, squares, and triangles are data for salts with monovalent, divalent, and trivalent counterions at $75~$mM, respectively. The inverse Debye's lengths of simulations are $\kappa^{-1} = 0.09\AA^{-1}$ for circles, $\kappa^{-1} = 0.11\AA^{-1}$ for squares and $\kappa^{-1} = 0.13\AA^{-1}$ for triangles. The lines are guides to the eyes.}
\label{fig5}
\end{figure}

We next explore the effect of the dielectric discontinuity across the  interface on the interaction between trapped colloidal particles.
In Fig.~\ref{fig5}, we show a comparison between the forces when the image charges are taken into account and when they are neglected. The induced charge at the air-water interface does not seem to affect significantly the interaction between the colloidal particles, except at very short separations.

\section{Conclusions}

We studied the interaction force  between colloidal particles trapped at an air-water interface.  The high electrostatic 
free energy penalty of exposing
charged groups to the low dielectric environment prevents ionization of these groups  inside either air or oil.  Therefore, only the surface groups which are hydrated will be ionized.  For these groups, ionization is favored by 
the entropic free energy gain of counterion release, while the enthalpic electrostatic free energy penalty is lowered by screening of the exposed surface charge groups by the dipole moments of the surrounding water molecules. In view of this, we have idealized the interface-trapped colloidal particles as hard-spheres with the hemisphere exposed to water carrying a uniform surface charge, while the other hemisphere
remaining charge neutral. We found that the
interface bound colloidal particles in 1:1 electrolyte 
interact with a force which is almost identical to the force  between bulk colloids of the same total charge.  The far-field weak dipole-dipole-like interaction can not be seen in our simulations. 
The equality between bulk and surface forces breaks down for colloids in 2:1 and 3:1 electrolytes, for which the bulk interaction is always more repulsive
than at the interface.  Finally, we have explored the effect of dielectric 
discontinuity across the interface on the inter-colloidal interaction.
For all the cases studied, 
we have found only a small effect of image charges at very short separations between the colloidal particles.  We hope that the work will provide a further stimulus to the development of analytical methods for studying colloidal particles trapped at dielectric interfaces between polar and non-polar mediums. 

\section{Acknowledgments}
This work was partially supported by the CNPq, INCT-FCx, and by the US-AFOSR under the grant 
FA9550-12-1-0438.

\bibliography{ref}

\end{document}